# Phase transitions in typical fluorite-type ferroelectrics

Heng Yu,[1] Kan-Hao Xue,[1,2,*] Nan Feng,[3] Yunzhe Zheng,[4] Yan Cheng,[4] Ben Xu,[3,*] and Xiangshui Miao[1,2]

[1]School of Integrated Circuits, Huazhong University of Science and Technology, Wuhan 430074, China

[2]Hubei Yangtze Memory Laboratories, Wuhan 430205, China

[3]Graduate School, China Academy of Engineering Physics, Beijing 100193, China

[4]Key Laboratory of Polar Materials and Devices (MOE), Department of Electronics, East China Normal University, 500 Dongchuan Road, 200241 Shanghai, China

***Corresponding Authors**, E-mail: xkh@hust.edu.cn (K.-H. Xue), bxu@gscaep.ac.cn (B. Xu)

## ABSTRACT

While ferroelectric hafnia ($HfO_2$) has become a technically important material for microelectronics, the physical origin of its ferroelectricity remains poorly understood. The tetragonal $P4_2/nmc$ phase is commonly assigned as its paraelectric mother phase but has no soft mode at the Brillouin zone center. In this work, we propose that the paraelectric—ferroelectric transition in the fluorite-type $Pca2_1$ ferroelectric family can be described by a $Pcca$—$Pca2_1$ transition, where the $Pcca$ mother phase will evolve into either the $Pca2_1$ ferroelectric phase or the centrosymmetric $P2_1/c$ monoclinic phase, depending on the strain conditions. The $Pcca$ phase is directly linked to both phases in the context of continuous phase transition. Hafnia is regarded as a special case of this family in that it has accidental atomic degeneracy because all anions are oxygen. The theory is also correlated to the seven-coordination theory that explains the ferroelectricity in hafnia from a chemical perspective. In addition, the strain conditions to promote the ferroelectric phase in hafnia are discussed.

Ferroelectricity in hafnia (HfO₂) and related materials have attracted enormous attention in recent years since a substantial spontaneous polarization could be maintained in these microelectronics-compatible dielectrics at nanometer thickness [1–3]. The physical understanding of such ferroelectricity, on the other hand, is still less satisfactory. Recently, we found that the special cation/anion radius ratio renders hafnia a seven-coordination (7C) configuration, which allows for extra room for some oxygen anions to alter their positions. In this sense, similar fluorite-type $Pca2_1$ ferroelectric candidates may be predicted, including SrI₂, YOF and LuOF [4,5]. This theory attacks the origin of ferroelectric hafnia from the aspect of "why", but one still needs to answer "how" hafnia becomes ferroelectric. While the ferroelectric phase is commonly ascribed to a $Pca2_1$ polar phase [6,7], its paraelectric mother phase and the zone center soft mode that should yield the desired paraelectric—ferroelectric transition are still uncertain [8].

In 2020, Lee and coworkers discovered that the ferroelectricity in hafnia, especially in terms of its high coercive field, is related to flat phonon modes [9]. It is shown that two modes $\Gamma_{15}^z$ and $Y_5^z$ from the same flat phonon band in cubic fluorite ($Fm\bar{3}m$) hafnia are relevant to the distortion into the ferroelectric $Pca2_1$ phase. Nevertheless, cubic $Fm\bar{3}m$ corresponds to a rather high temperature phase of hafnia, and its apparent soft mode drives the structure into the tetragonal $P4_2/nmc$ phase. Most researchers regard the tetragonal $P4_2/nmc$ phase as the mother phase [10–12]. Using such a paraelectric phase, Delodovici et al. systematically studied the transition from $P4_2/nmc$ to $Pca2_1$ [13]. Upon freezing the $\Gamma_{4+}$ mode, the structure could enter a $Ccce$ space group, which only involves one soft mode, whose impact is to drive the structure into an $Aea2$ phase nevertheless. It is only through a synergic effect of the $\Gamma_{3-}$ ($Aea2$), $Y_{2+}$ ($Pbcn$), and $Y_{4-}$ ($Pcca$) modes that the ferroelectric $Pca2_1$ phase could be obtained. Moreover, the effect of strain plays a significant role [14,15]. Zhou et al. suggested that tensile strain softens the antipolar mode, causing a transition from $P4_2/nmc$ to $Pbcn$. Subsequently, the polar-antipolar coupling stabilizes the ferroelectric $Pca2_1$ phase [16].

Even though such a three-phonon coupling process could explain the ferroelectric transition in hafnia, treating $P4_2/nmc$ as the mother phase of ferroelectric hafnia is still unsatisfactory in two aspects. On the one hand, it does not follow the traditional scenario that a single zone-center mode renders a paraelectric-ferroelectric transition [8]. On the other hand, both $Pca2_1$ hafnia and the ground state $P2_1/c$ hafnia are in 7C configuration, while the high temperature phases (cubic fluorite, tetragonal $P4_2/nmc$) show eight-coordination (8C) for the cations. Upon cooling, the 8C—7C transition could yield either the $P2_1/c$ phase (symmetric transition) or the $Pca2_1$ phase (asymmetric transition), depending on the strain situation. An ideal mother phase should generate either phase in one step, upon specifying the environmental circumstances. In this work, we propose that a $Pcca$ phase is such a proper paraelectric mother phase for hafnia and similar fluorite-type $Pca2_1$ ferroelectric candidates. Furthermore, the strain condition will be specified to determine the direction of transition.

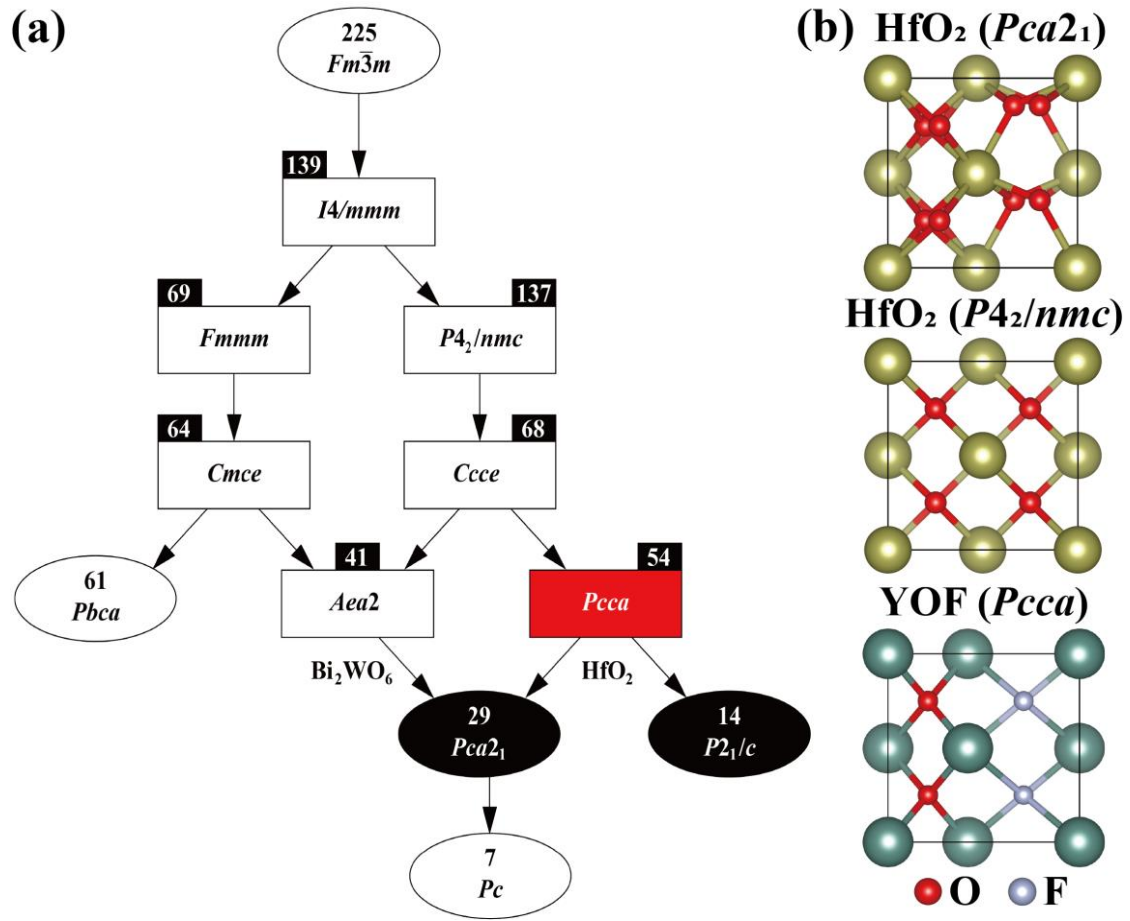

**Figure 1.** (a) The continuous transition ways from the $Fm\bar{3}m$ phase towards $Pca2_1$ and $P2_1/c$. (b) Crystal structures of orthorhombic $Pca2_1$ HfO₂, tetragonal $P4_2/nmc$ HfO₂ and orthorhombic $Pcca$ YOF. Hafnium, oxygen, yttrium and fluorine atoms are in gold, red, green and gray colors, respectively.

A symmetry analysis based on the $Fm\bar{3}m$ fluorite phase serves as the starting point [17,18]. The continuous transition ways towards $Pca2_1$ and $P2_1/c$ are illustrated in **Figure 1**. Besides hafnia, another well-known ferroelectric material Bi$_2$WO$_6$, in the $Pca2_1$ phase as well, is selected for comparative reasons [19]. $Fm\bar{3}m$ is the highest symmetry phase of hafnia in its solid-state form, while the high temperature paraelectric phase of Bi$_2$WO$_6$ is tetragonal $I4/mmm$, which is a subgroup of $Fm\bar{3}m$. Of course, the layered structure of Bi$_2$WO$_6$ forbids any cubic symmetry. To obtain the ferroelectric phase, $I4/mmm$ is thus a common starting point for both materials. As a member of the Aurivillius layered-perovskite family, the transition to the ferroelectric Bi$_2$WO$_6$ phase involves $I4/mmm \rightarrow Fmmm \rightarrow Aea2 \rightarrow Pca2_1$ [20]. A slight lattice constant discrepancy involves the first transition to $Fmmm$, which further becomes ferroelectric $Aea2$ through the mutual effect of an $Fmm2$ mode and a $Cmce$ (which was also named $Bmab$ in old references, as a non-standard form of $Cmca$, but $Cmca$ has been modified as $Cmce$ since 1995) mode [21]. A similar Aurivillius-phase ferroelectric, Bi$_4$Ti$_3$O$_{12}$, was also regarded as in an $Aea2$ space group (usually named $B2cb$ in old references), though its non-vanishing polarization component along $c$-axis implies that its true ground state symmetry is $Pc$ ($B1a1$) [22], which comes from $Aea2$ through a $Pca2_1$ intermediate [23]. Hence, the $Pca2_1$ structure of Bi$_2$WO$_6$ may be regarded as a distorted version of the $Aea2$ structure, since it merely loses the base-centered orthorhombic symmetry element.

The transition process in hafnia, however, is quite distinct from Bi$_2$WO$_6$ because it definitely comes through a $P4_2/nmc$ intermediate phase, as is well-established from experiments [6]. As pointed out by Delodovici et al., symmetry broken in hafnia further involves a $Ccce$ mode, but three modes exist for $Ccce \rightarrow Pca2_1$. Among the three modes ($Aea2$, $Pbcn$, $Pcca$), the two modes that could yield both $Pca2_1$ and $P2_1/c$ phases are $Pbcn$ and $Pcca$. It is indeed tempting to regard the $Pbcn$ or $Pcca$ phase as the paraelectric mother phase of hafnia, for (i) it yields the ferroelectric phase through softening a single zone-center mode only; (ii) it is the direct mother phase for both $Pca2_1$ and $P2_1/c$. Nevertheless, no $Pbcn$ or $Pcca$ hafnia has been reported yet. This forces us to enlarge our scope of materials in order to extract the relevant physics.

Based on the 7C theory, several ternary ferroelectric candidates in the $Pca2_1$ structure have been predicted, including YOF, LuOF, and TaON. The methodology is to substitute a cation other than +4 valency for Hf, while replacing half of the anions correspondingly. Indeed, the O anions with three-coordination (referred to as $O_{III}$) and with four-coordination (referred to as $O_{IV}$) are distinct in $Pca2_1$-$HfO_2$, and doing such elemental substitution does not degrade its space-group symmetry. The symmetric 7C form of these ternary compounds was found to be $P2_1/c$, similar to hafnia. Suppose one attempts to obtain the paraelectric mother phase of these ternary materials, one immediately realizes that any simple tetragonal phase is impossible. The resulting mother phase, established based on a hypothetical "$P4_2/nmc$" structure, has a $Pcca$ symmetry. Hence, the paraelectric-ferroelectric transition in these compounds is simply $Pcca \rightarrow Pca2_1$. In hafnia, all O anions are $O_{IV}$ in the $P4_2/nmc$ phase, but any of its potential 7C phase must involve both $O_{III}$ and $O_{IV}$. If this distinction may occur before the tetragonal→orthorhombic transition, then the space group symmetry of hafnia should directly enter $Pcca$, without the need of invoking a $Ccce$ intermediate.

To confirm or criticize the above argument, we attempted to perturb the O anions in $P4_2/nmc$ hafnia using various techniques. The first method is to introduce some differences in the atomic potential between the ultimate $O_{III}$ and $O_{IV}$ anions or in the local parts of their pseudopotentials. Hence, we adopted a DFT-1/2 method [24–26], which involves attaching a self-energy potential for the anions, but in our case, this extra potential is multiplied by a small factor +A and -A, unequally to two sorts of O anions [27]. The second method is orbital-dependent DFT+U [28], where we impose tiny positive U-J and negative U-J values to the two sorts of O anions. In either case, the structure immediately enters a $Pcca$ symmetry, even if A is as small as 0.01 or U is merely 0.1 eV. After structural relaxation, the lattice slightly distorts into an orthorhombic structure, but the $Pcca$ symmetry is preserved. We then gradually increased the value of A and U, while the total energy of the relaxed structure continued to decrease due to symmetry broken, but the space group is always $Pcca$. Hence, one may conveniently study the impact of strains and other parameters based on a $Pcca$ phase, to determine the situation when symmetric distortion into $P2_1/c$ or asymmetric distortion into $Pca2_1$ is favored. Moreover, it is permitted to calculate the phonon spectra based on fully relaxed *Pcca*-hafnia.

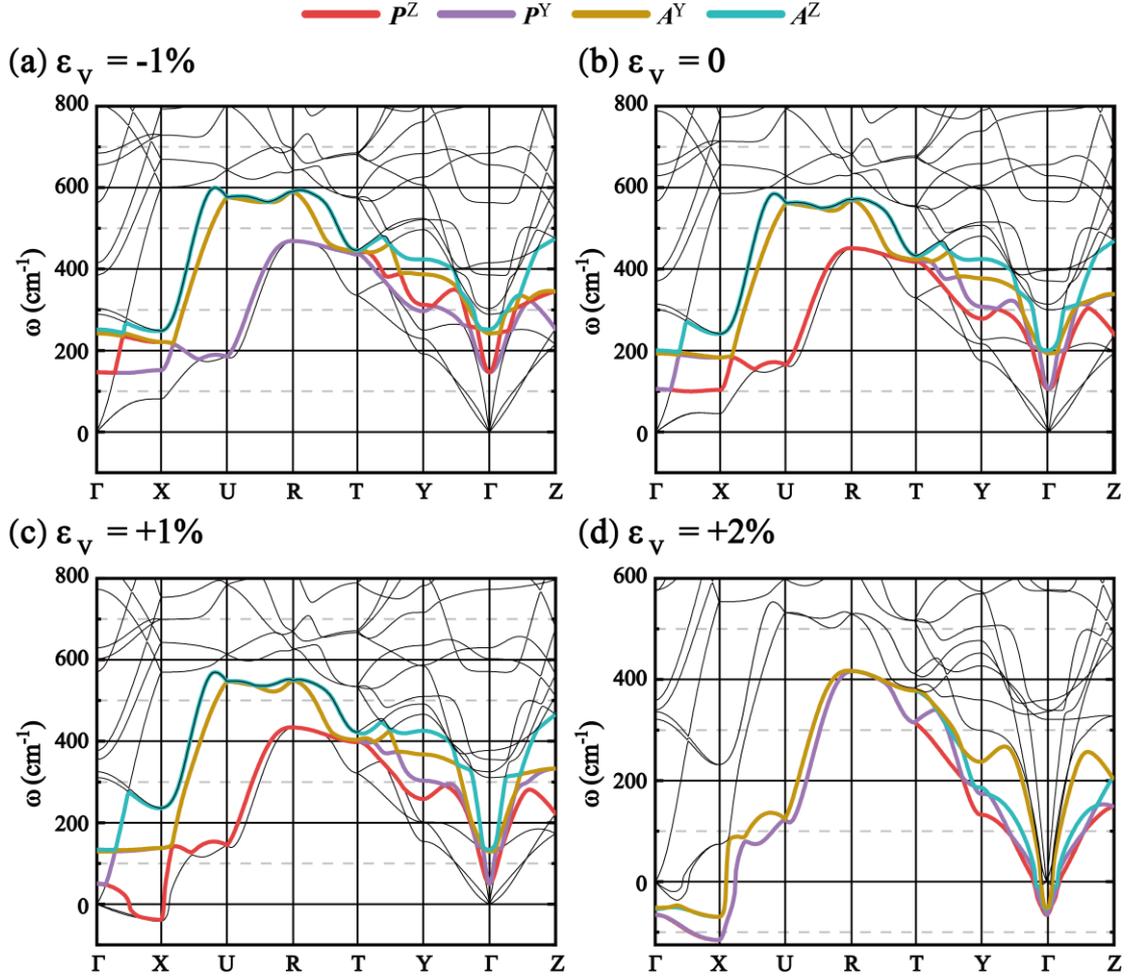

**Figure 2.** The phonon spectra of *Pcca* phase hafnia under various volume-strains. (a) Under compressive strain, the frequencies of the four phonon modes increase. (b) Without strain, no imaginary frequency exists in the phonon spectrum. (c) Under 1% tensile strain, the frequencies of the four phonon modes decrease rapidly. (d) Under 2% tensile strain, more severe imaginary frequencies are observed in these four phonon modes.

**Figure 2(b)** illustrates the phonon spectra of *Pcca* hafnia obtained through DFT+A-1/2 (A=0.01), which does not exhibit any imaginary mode. As is well known, substrate-induced strain makes a significant contribution to the stability of ferroelectricity in hafnia. The simplest strain configuration is related to volume expansion or contraction, with equal strains assigned along the three primitive vectors. A volume-strain parameter $\varepsilon_V$ is introduced such that

$$V' = (1 + \varepsilon_V)V_0 ,$$

where $V_0$ is the strain-free ground state volume, and $V'$ is the volume under strain. The cubic, tetragonal ($P4_2/nmc$) as well as $Pcca$ structures of hafnia, though different in cell volumes, are all relatively compact phases compared with the monoclinic $P2_1/c$ and ferroelectric $Pca2_1$ structures. Hence, $\varepsilon_V$ can be slightly negative but may be considerably positive, in order to meet practical situations.

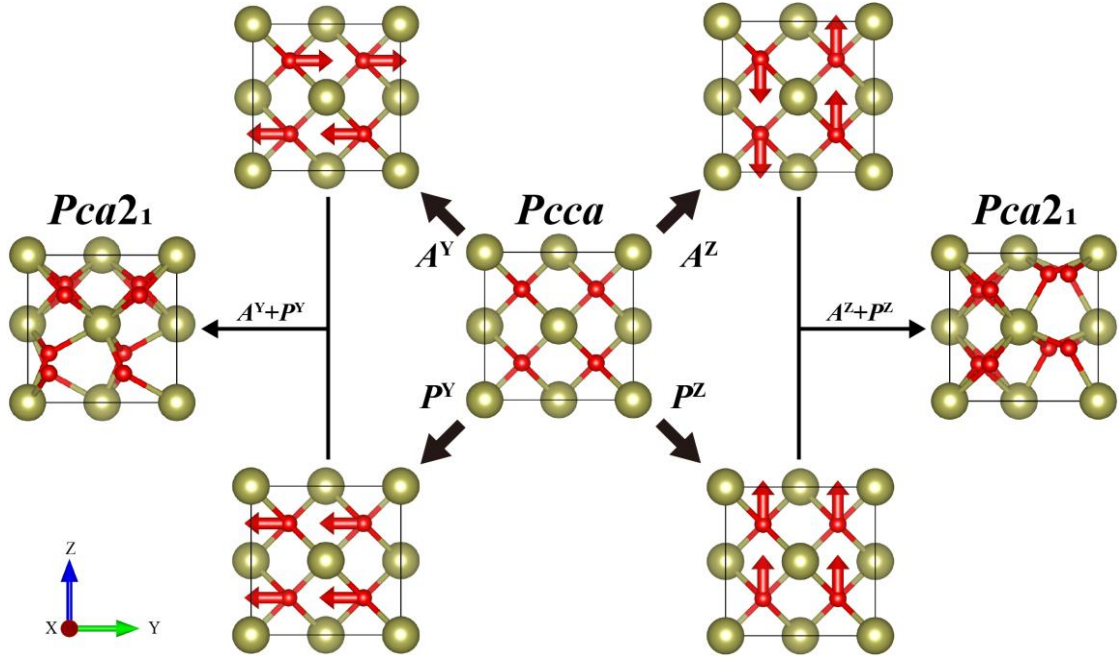

**Figure 3.** Oxygen displacement of four phonon modes in $Pcca$ HfO$_2$. Big gold balls and small red balls indicated hafnium and oxygen, respectively. Red arrows denote the oxygen atom displacements.

Several $\varepsilon_V$ values have been considered in this work, including -1% (compressive), 0%, 1% (tensile), and 2% (tensile). Under the compressive strain, all phonon modes of the $Pcca$ phase are stable. However, under tensile strains, the phonon dispersions develop instabilities as shown in **Figure 2**. Four phonon modes are weakly unstable at the Brillouin zone center, labelled as $P^Z, P^Y, A^Y, A^Z$. The $P$ and $A$ denote the polar mode and antipolar mode, respectively. The superscript denotes the direction of atomic displacement. The four phonon modes are condensed during the transformation from $Pcca$ to $Pca2_1$. The $P^Z$ mode in which all oxygen atoms move along the Z direction results in a uniform polarization. Regarding the $A^Z$ mode, oxygen atoms in adjacent XZ planes move along the Z axis in an antiparallel manner. The atomic motions under $P^Y$ and $P^Z$, as well as $A^Y$ and $A^Z$ are similar, but $P^Y$ and $A^Y$ are related to atomic movements along the Y direction. It then follows that $P^Z$ and $A^Z$ can yield a $Pca2_1$ structure polarized along the original Z direction, while $P^Y$ and $A^Y$ can yield another $Pca2_1$ structure polarized along the original Y direction, as illustrated in **Figure 3**. The formation of the ferroelectric $Pca2_1$-HfO$_2$ phase is closely related to the above four modes.

In comparison to $P4_2/nmc$, $Pcca$ has two advantages as the parent paraelectric phase for the phase transition mechanism in ferroelectric hafnia. On the one hand, the high symmetry $P4_2/nmc$ is not robust against even tiny strains and typically degrades to a $Ccce$ space group. However, $Pcca$ well maintains its space group characteristic under ordinary stress applied in any direction. On the other hand, with external pressure, the $Pcca$ phase exhibits all imaginary modes corresponding to the ferroelectric phase transition, while $P4_2/nmc$ only exhibits $P^Y$ and $P^Z$ at the Brillouin zone center under stress (**Figure S1**).

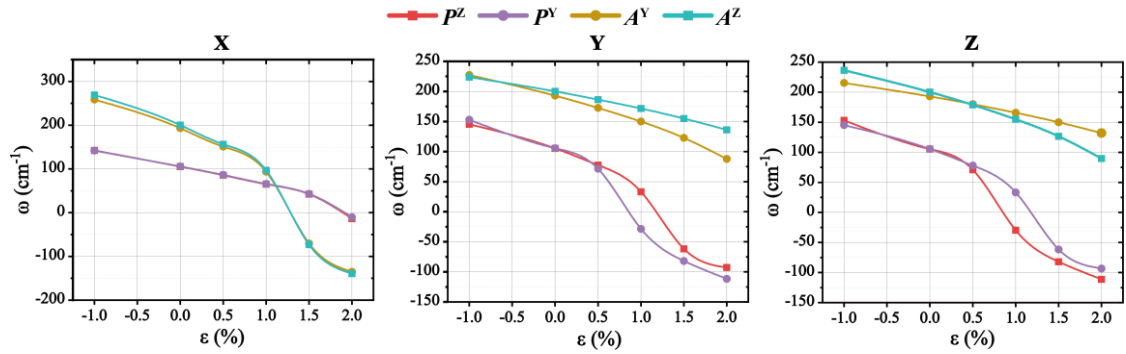

**Figure 4.** Variations in the frequency of the four phonon modes with respect to the strain applied to each direction.

According to their frequency variation behaviors under various strain conditions, the phonon modes that are closely related to the ferroelectric phase transition can be identified. Furthermore, to better investigate the effects of strain in different directions on the frequencies of these four phonon bands, we applied strain ε in each particular direction, with values ranging from -1% to +2% in **Figure 4**. When the strain is applied in the X direction, the magnitudes of the frequency changes for the $P^Y$ and $P^Z$ modes are the same, as well as for $A^Y$ and $A^Z$. However, the magnitudes of change in $A^Y$ and $A^Z$ are noticeably greater than that of $P^Y$ and $P^Z$. Therefore, external strains applied in the X direction primarily soften $A^Y$ and $A^Z$. A tensile strain in the Y direction separates $P^Y$ and $P^Z$, as well as $A^Y$ and $A^Z$, and mainly affects $P^Y$. However, when the tensile strain in the Y direction becomes substantial, such as ε=+2%, it shifts $P^Y$ and $P^Z$ almost by an equal extent. Similarly, tensile strain in the Z direction separates $P^Y$ and $P^Z$, as well as $A^Y$ and $A^Z$, and mainly affects $P^Z$. To sum up, tensile strain in the X direction primarily softens $A^Y$ and $A^Z$, while tensile strain in the Y and Z direction respectively soften $P^Y$ and $P^Z$. The same treatment was applied to ZrO$_2$, leading to similar conclusions as shown in **Figure S3**. However, the four phonon modes of ZrO$_2$ are more stable under stress. Significant changes in the phonon mode frequencies were only observed then when the value of ε reached 2%. According to seven-coordination theory, this is due to the larger ρ (ρ = $r_{cation}$/$r_{anion}$) in ZrO$_2$ compared to HfO$_2$, which results in a relatively smaller space for oxygen atoms to move in ZrO$_2$.

Hence, it is natural to expect that the external condition to stabilize the $Pca2_1$ phase is to stretch its *a*-direction while keeping its *b*-direction unstretched or even compressed. Following Zhou et al. [16], an aspect ratio parameter can be defined as

$$R_a = \frac{2a}{c+b}.$$

An experimental measurement of the $Pca2_1$ phase of hafnia reveals that $R_a$ is relatively large, typically greater than 1.1 [29]. Nevertheless, calculation result on an ideal $Pca2_1$ phase of hafnia gives an $R_a$ value of merely 1.04. Following the examples given above, our predicted favorable strain conditions for ferroelectric hafnia indeed tend to increase the orthorhombic aspect ratio $R_a$. This implies that special substrate (and/or top electrode interfacial) conditions ought to be imposed to stabilize the $Pca2_1$ phase of hafnia, the features of which agree well with our analysis based on the phonon spectra of $Pcca$ hafnia.

A potential theoretical difficulty in assigning the $Pcca$ phase as the paraelectric mother phase of hafnia lies in that, all O anions are identical if viewed from the $P4_2/nmc$ phase. Nevertheless, the ultimate structure at room temperature for hafnia usually discriminates the O$_{III}$ anions from the O$_{IV}$ anions. Any stress conditions could easily destroy the atomic symmetry between the O anions in hafnia. Although we have to manually attach some asymmetric gradients to hafnia to reach the $Pcca$ phase above, this can become natural if the dielectric is placed in a designed environmental situation such that the ferroelectric phase may become stable. Hence, we have merely triggered this process without resorting to the complicated environment in this research. Moreover, if one regards hafnia as a special member of this $Pca2_1$ ferroelectric family, it follows easily that $Pcca$ is the generic paraelectric mother phase, and hafnia merely has some atomic degeneracy. Actually, provided that the Y-coordinates of the Hf cations are unequally distributed within a supposed $P4_2/nmc$ unit cell, it will render the O anions different in chemical environment, transforming the $P4_2/nmc$ symmetry into $Pcca$ symmetry (see **Figure S5**).

For a scrutiny into another member of this ferroelectric family, **Figure S6** depicts the phonon dispersion curves for the $Pcca$ phase of YOF. At the Γ point, a soft mode with symmetry $\Gamma_4^-$ captures the distortion connecting the $Pcca$ and a new $Pba2$ structure. As the volume expands, $\Gamma_4^-$ becomes stable, while another phonon mode $\Gamma_3^-$ loses its stability. The $\Gamma_3^-$ mode leads to the orthorhombic $Pca2_1$ phase. Tensile stress along the X or Y directions will stabilize $\Gamma_4^-$, whereas tensile stress applied in any direction will transform the $\Gamma_3^-$ mode into a soft mode. Therefore, YOF is susceptible to transforming into the ferroelectric $Pca2_1$ phase. Unlike HfO$_2$, transition from $Pcca$ to $Pca2_1$ in YOF results in a unique polarization direction, where the F atoms only move along the original Z direction.

In conclusion, we have assigned a new paraelectric mother phase ($Pcca$) for the generic family of fluorite-type $Pca2_1$ ferroelectric materials, and investigated the impact of strain on the stabilization of the ferroelectric phase from the phonon spectra perspective. Under compressive strains, all phonon modes in the $Pcca$ structure are stable for hafnia. Tensile strain in the X direction primarily drives the phase transition from $Pcca$ to $Pca2_1$, while tensile strain in the Y or Z direction determines the potential polarization direction of $Pca2_1$. The $Pcca$- $Pca2_1$ transition servers as the prototype of the paraelectric-ferroelectric transition in this family of ferroelectric candidates, and hafnia is special in that it has an accidental atomic degeneracy between the 3-coordination and 4-coordination anion sites.

# Acknowledgements

This work was supported by the National Key R&D Program of China under Grant No.

Supplementary Material

# Phase transitions in typical fluorite-type ferroelectrics

Heng Yu,[1] Kan-Hao Xue,[1,2,*] Nan Feng,[3] Yunzhe Zheng,[4] Yan Cheng,[4] Ben Xu,[3,*] and Xiangshui Miao[1,2]

[1]School of Integrated Circuits, Huazhong University of Science and Technology, Wuhan 430074, China

[2]Hubei Yangtze Memory Laboratories, Wuhan 430205, China

[3]Graduate School, China Academy of Engineering Physics, Beijing 100193, China

[4]Key Laboratory of Polar Materials and Devices (MOE), Department of Electronics, East China Normal University, 500 Dongchuan Road, 200241 Shanghai, China

*Corresponding Authors, E-mail: xkh@hust.edu.cn (K.-H. Xue), bxu@gscaep.ac.cn (B. Xu)

# Computational settings

Density functional theory (DFT) calculations were performed within the Vienna ab initio simulation package (VASP) using the projector-augmented wave (PAW) method [1,2]. The valence electron configurations were 2s and 2p for O/F; 4s, 4p, 4d, and 5s for Y/Zr; 5p, 5d, and 6s for Hf. The Perdew-Burke-Ernzerhof functional generalized gradient-approximation type (GGA-PBE) was adopted to treat the exchange-correlation [3]. A cutoff energy of 600 eV and a $10 \times 10 \times 10$ Monkhorst Pack k-mesh were used in the calculations. The atomic positions and lattice constants were fully relaxed until the atomic Feynman forces were smaller than 0.001 eV/Å in each direction. The force constants are calculated using a 2×2×2 supercell and a 4×4×4 k-point grid. The phonon spectrum at zero temperature is then computed using the finite difference method and postprocessed in the PHONOPY code [4].

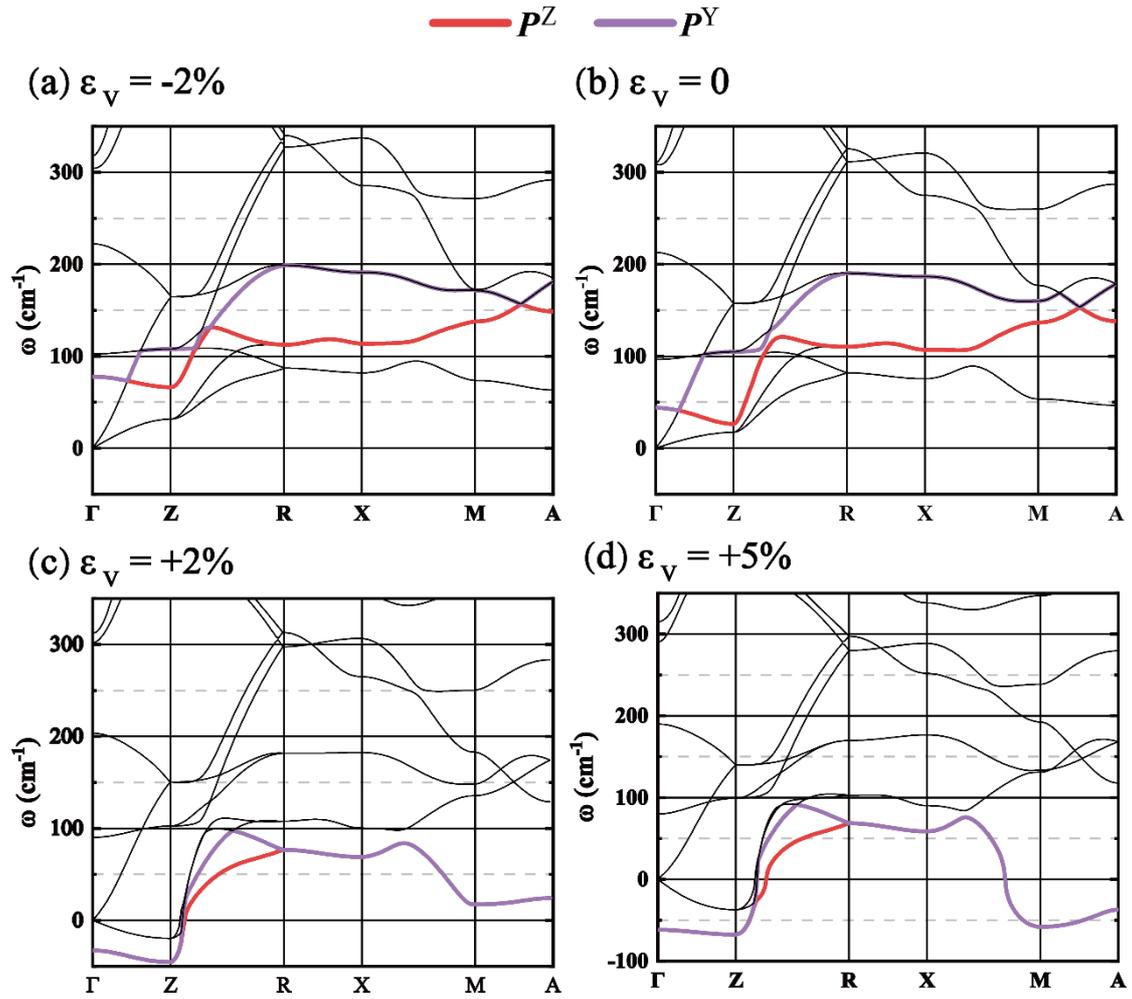

**Figure S1. The phonon spectra of tetragonal $P4_2/nmc$ hafnia under various volume-strains.** (a) Without strain, no imaginary frequency exists in the phonon spectrum. (b) Under 2% tensile strain, the frequencies of the two polar phonon modes decrease rapidly. (c) Under 5% tensile strain, more severe imaginary frequencies are observed in the two polar phonon modes. However, there are no other imaginary frequencies present.

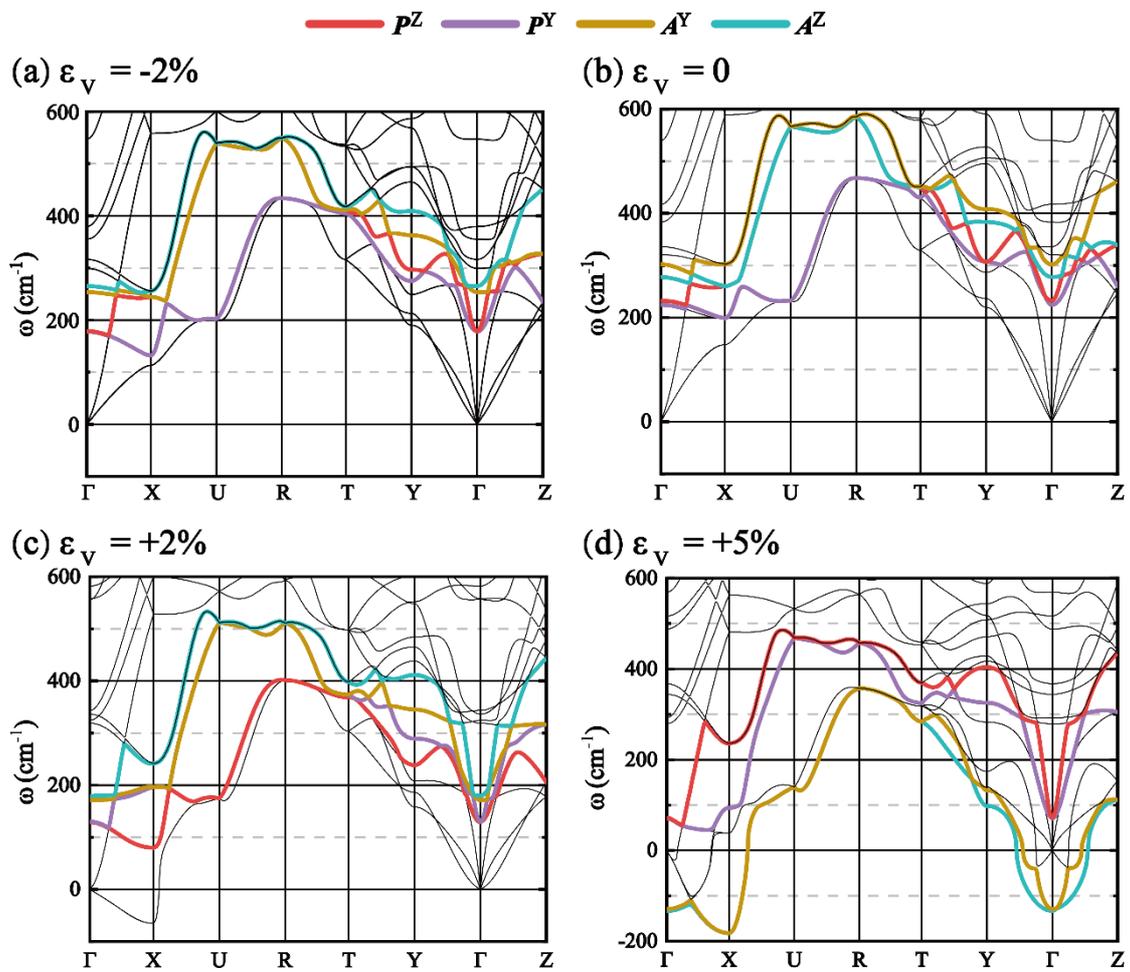

**Figure S2. The phonon spectra of *Pcca* phase zirconia under various volume-strains.** (a) Under compressive strain, the frequencies of four phonon modes increase. (b) Without strain, no imaginary frequency exists in the phonon spectrum. (c) Under 2% tensile strain, the frequencies of the four phonon modes decrease rapidly. (d) Under 5% tensile strain, more severe imaginary frequencies are observed in the two antipolar modes.

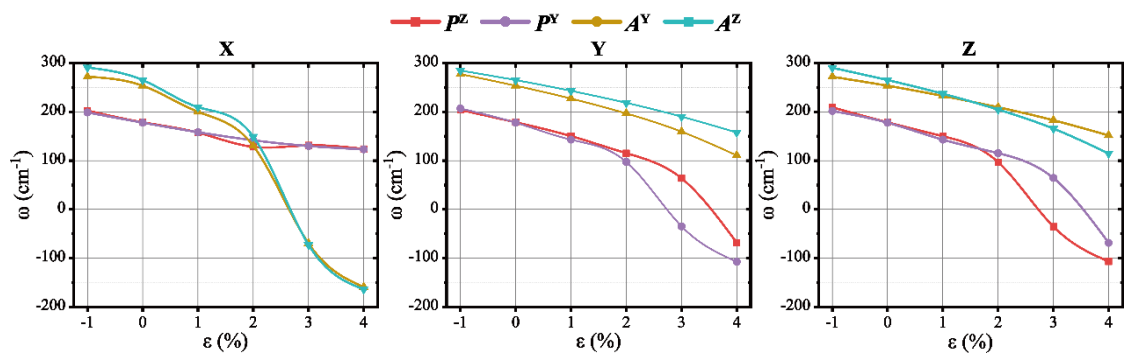

**Figure S3. The frequency variations of four phonon modes in zirconia with respect to the strain applied to each direction.**

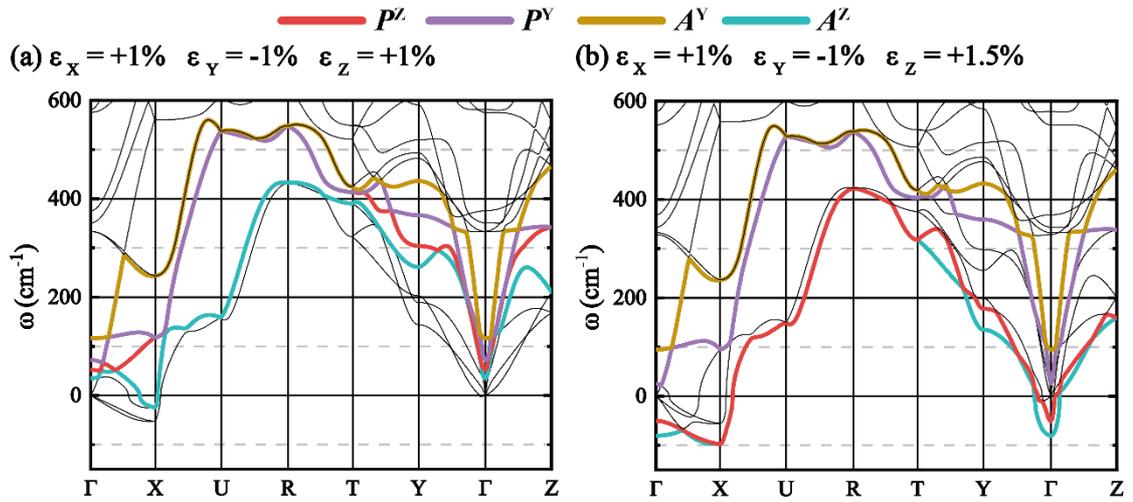

**Figure S4. The phonon spectra of *Pcca* hafnia under specific strain conditions.**

Suppose one aims at obtaining a $Pca2_1$ phase hafnia with its polarization aligned in the Z direction of the original *Pcca* phase, tensile strains can be applied in the X and Z directions to soften $P^Z$ and $A^Z$. In the meantime, one needs to apply a compressive strain in the Y direction to stabilize $P^Y$ and $A^Y$. Therefore, we first attempt to apply strains of $\varepsilon_X$=+1%, $\varepsilon_Y$=-1%, and $\varepsilon_Z$=+1% to *Pcca* hafnia. The phonon spectra show that while $P^Z$ and $A^Z$ get softened, they have not become imaginary frequencies yet. To achieve the phase transition from *Pcca* to $Pca2_1$ with polarization along the original Z direction, we continue to apply tensile strain in the Z direction, i.e., $\varepsilon_Z$=+1.5%. Under this condition, $P^Z$ and $A^Z$ become imaginary, driving the phase transition.

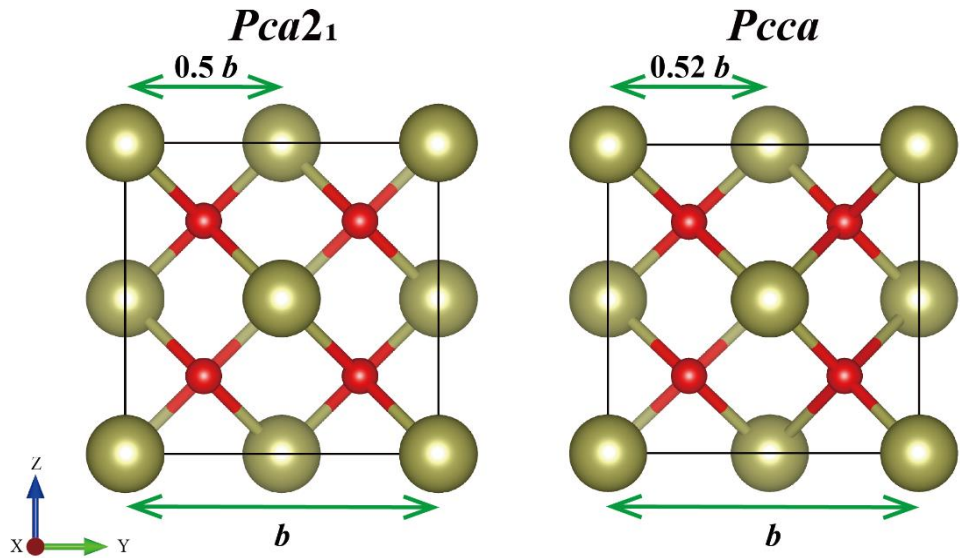

**Figure S5. The space group transition in tetragonal $P4_2/nmc$ hafnia due to the unequal distribution of Hf cations in the Y direction.**

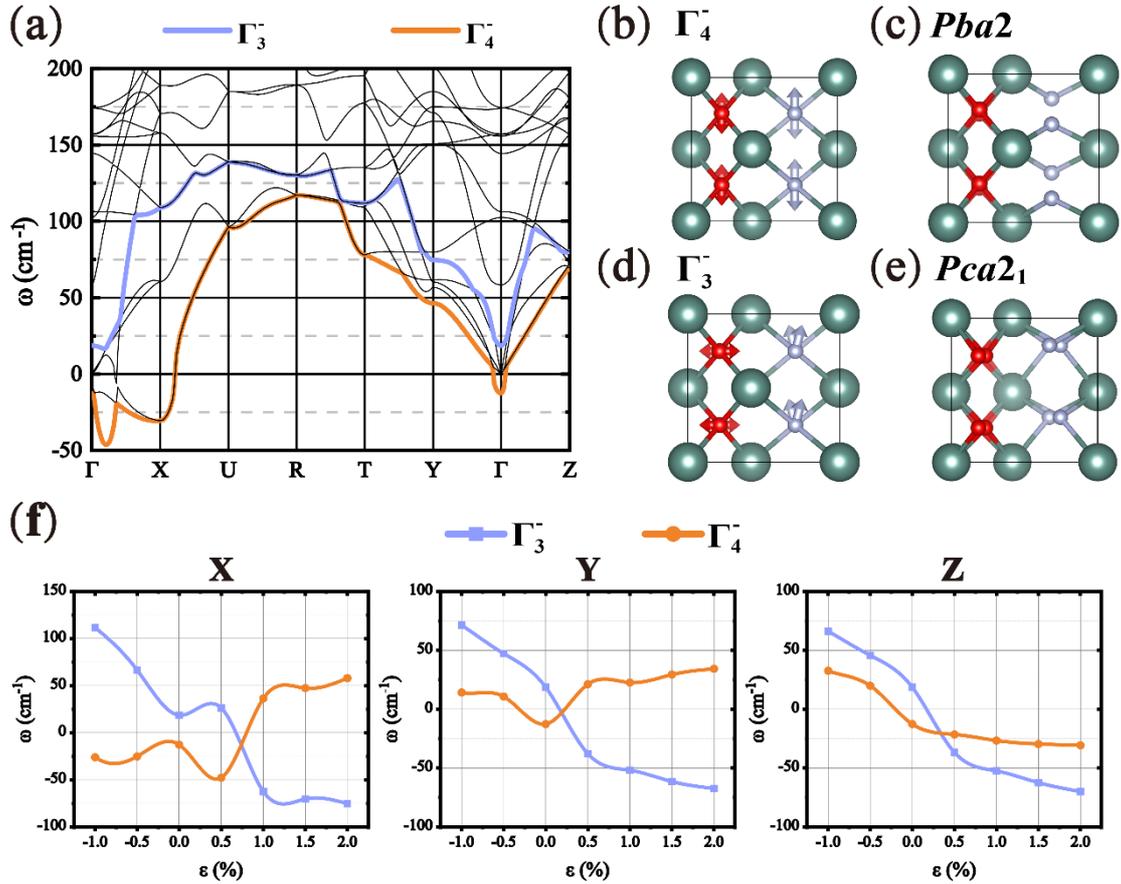

**Figure S6.** (a) Phonon spectrum of the YOF *Pcca* phase, where the most notable phonon modes are in bold format. (b) The atomic motions of $\Gamma_4^-$ mode; (c) The *Pba*2 phase corresponding to the $\Gamma_4^-$ mode. (d) The atomic motions of $\Gamma_3^-$ mode; (e) The *Pca*$2_1$ phase corresponding to the $\Gamma_3^-$ mode. (f) Variations in the frequency of the two phonon modes with respect to the strain applied to each direction.